Regular Article - Experimental Physics

# Optimum filter synthesis with DPLMS method for energy reconstruction

V. D'Andrea[1,5,a] , S. Riboldi[2], A. Geraci[3], N. Burlac[4], F. Salamida[1]

[1] Dipartimento di Scienze Fisiche e Chimiche, Università degli Studi dell'Aquila, L'Aquila and INFN LNGS, Assergi, Italy
[2] Dipartimento di Fisica, Università degli Studi di Milano and INFN Milano, Milan, Italy
[3] Dipartimento di Elettronica, Informazione e Bioingegneria, Politecnico di Milano and INFN Milano, Milan, Italy
[4] Dipartimento di Matematica e Fisica, Università degli Studi Roma Tre and INFN Roma Tre, Rome, Italy
[5] *Present address:* INFN Roma Tre, Rome, Italy

**Abstract** Optimum filters are granted increasing recognition as valuable tools for parametric estimation in many scientific and technical fields. The DPLMS method, introduced some twenty years ago, is especially effective among the synthesis algorithms since it derives the optimum filters directly from the experimentally acquired signal and noise waveforms. Two new extensions of the DPLMS method are here presented. The first one speeds up the synthesis phase and improves the energy estimation results by synthesizing optimum filters with automatically designed flat-top length. The second one improves the quality of parameter estimation in multi-channel systems by taking advantage of the inter-channel noise correlation properties. In this paper, the theoretical and functional aspects behind the DPLMS method for optimum filter synthesis are first recalled and illustrated in more detail. The two new DPLMS extensions are subsequently briefly introduced from the theoretical viewpoint and more thoroughly considered from the applicative perspective. The DPLMS optimum filters have been applied first to simulated signals with various amounts and characteristics of superimposed noise and then to the experimental waveforms acquired from a solid-state Ge detector. The results obtained are considered from both the absolute viewpoint and in comparison with those of more traditional, suboptimal filters. The obtained results demonstrate the effectiveness of the two new DPLMS extensions. For single-channel energy estimations, the synthesized optimum filters provide comparatively better results than the other tested filters. The DPLMS multi-channel optimum filters further enhance the quality of the estimations, compared to single-channel optimum filters, with non-negligible inter-channel noise correlation. The effectiveness and robustness of the DPLMS method in synthesizing high-quality filters for energy estimation will be tested soon within leading-edge multi-channel physics experiments.

## 1 Introduction

The precise estimation of various parameters, such as energy and time from signals collected in a noisy environment, is usually accomplished by filtering techniques. The origin and characteristics of noise can be different from case to case in the various fields of application. Several optimum filter techniques to perform parameter estimation based on the characterization of the specific noise, already exist in the literature, e.g., the Wiener algorithm [1], the discrete Fourier transform method [2], and the pole identification method [3].

A general approach to this problem is provided by the class of Least Mean Squares (LMS) procedures [4,5], which calculate the optimum filter from the estimation of the experimental noise auto-correlation function. The Digital Penalized LMS (DPLMS) method, considered in the present work, is also based on the LMS approach and has already been proposed and applied in the past [6–8].

Despite their potential benefits, optimum digital filters have not yet typically spread to their possible fields of application. This is probably because of both the additional computational challenge of implementing adaptive filtering and the underestimation of their potential improvements in comparison to standard filtering techniques. Notable exceptions of modern physics experiments, relying on the DPLMS method for signal processing, include [9], based on Ge detectors, and [10,11], based on silicon drift detectors, both reporting significant improvements.

[a] e-mail: valerio.dandrea@roma3.infn.it (corresponding author)

Springer

This work will show the details of the optimum filter synthesis for energy estimation with the DPLMS method and its application to experimental setups. At the same time, we will demonstrate the technique's effectiveness in various noise configurations compared to commonly used digital filters.

The paper is organized as follows. In Sect. 2, starting from the basic concepts of the DPLMS method, we will report on a few recent improvements and briefly outline two new extensions of the original algorithm. In Sect. 3, we will report in detail the implementation procedure of the DPLMS algorithm. In Sect. 4, we will show several examples of DPLMS synthesized optimum filters and demonstrate their effectiveness in comparison to suboptimal filters in terms of energy estimation results. In Sect. 5, we will focus on designing DPLMS optimum filters for multi-channel applications, thanks to the second algorithm extension, and show a few examples of the synthesized filters with their corresponding results. The two new DPLMS extensions will be more comprehensively addressed in a forthcoming paper.

## 2 The DPLMS method

As a general concept, filter optimality is usually naively associated with optimal extraction of information of interest (e.g., the amount of deposited energy) out of a time-limited portion of detectors' signals, potentially affected by noise. However, that goal may be practically achieved with different procedures, each targeted for a specific application, which partially explains the many different results obtained in the optimum filter search.

Apart from the almost ubiquitous goal of noise reduction, the optimum filters must sometimes also cope with other potentially detrimental misestimation effects to achieve the overall best measurement results. Examples include the ballistic deficit in large volume detectors, the pulse pileup in the case of high rate acquisitions, or the microphonics and power-line disturbance pickup.

Consequently, the appropriate blending of desirable filter properties is always strictly application-dependent. Additionally, the conflicting nature of the goals usually prevents achieving an overall satisfying filter synthesis by satisfying just a single constraint (e.g., noise minimization).

The DPLMS approach to optimum filter synthesis intrinsically accomplishes the complex nature of the diverse applicative scenarios by associating a distinct weighting coefficient to each desired filter constraint. Thus automatically synthesizing filters that globally satisfy a weighted combination of the possibly conflicting goals.

As a result, the DPLMS approach automatically synthesizes filters with better overall properties compared to less automated methods (that require manual adjustments [4]). The characteristics signal traits, in terms of deterministic pulse shape and statistical properties, are derived from experimentally acquired waveforms. Therefore, the DPLMS synthesized filters can account for all noise components (e.g., narrow band disturbances, electromagnetic interference) in addition to the fundamental ones (e.g., series and parallel noise in the case of radiation detectors). It can also reject the pulse pileup and ballistic deficit conditions.

We now provide, in the following, a more comprehensive description of the DPLMS algorithm, compared to the first publication [7] and add two new algorithm extensions.

Estimating a single parameters (such as energy or time) out of the many-sample signal waveforms is a data reduction process performed by a digital filter. Since the DPLMS method performs the optimum filter search within the class of the finite impulse response (FIR) linear filters, the filter output signal is the linear combination of the input signal waveform $\psi_{in}$ weighted by the filter coefficients.

Any dataset of signals $\psi_{in}$, i.e., those composed of filter input waveforms, can always be associated with a deterministic and a non-deterministic component:

$$\overline{\psi}_{in} = E[\psi_{in}] \tag{1}$$

and

$$Var[\psi_{in}] = E[\psi_{in} - \overline{\psi}_{in}]^2. \tag{2}$$

In the simplest case of detector signals with constant shape, Eq. (1) would represent the noiseless detector response signal and Eq. (2) the first-order statistical description of the superimposed noise.

The output response of a linear system is the time convolution of the filter input signal $\psi_{in}$ and the filter impulse response $h$:

$$\psi_{out}(n) = h * \psi_{in} = \sum_{i=-\infty}^{+\infty} h(i) \cdot \psi_{in}(n-i). \tag{3}$$

In the case of FIR filters with limited-length non-zero inpulse response $h(n)$ (for $0 \leq n < N$), Eq. (3) simplifies to:

$$\psi_{out}(n) = \sum_{i=0}^{N-1} h(i) \cdot \psi_{in}(n-i). \tag{4}$$

The same result can be equivalently expressed in terms of the filter weighting function $w(i) = h(-i)$:

$$\psi_{out}(n) = \sum_{i=0}^{N-1} w(-i) \cdot \psi_{in}(n-i) \tag{5}$$

or in terms of the optimum filter coefficients:

$$x = [x_1, x_2, ..., x_N] \quad (6)$$

(defined so that $x_i = w(i - N)$):

$$\psi_{out}(n) = \sum_{i=1}^{N} x_i \cdot \psi_{in}(n - N + i). \quad (7)$$

By comparing Eqs. (3) and (7), the impulse response $h$ of the optimum filter relates to the coefficients $x$ as:

$$h(i) = x_{N-i}. \quad (8)$$

Given the filter input signals dataset, $\psi_{in}$, the deterministic and non-deterministic components of the filter output response can be expressed respectively as:

$$\overline{\psi}_{out}(n) = h * \overline{\psi}_{in} = \sum_{i=1}^{N} x_i \cdot \overline{\psi}_{in}(n - N + i) \quad (9)$$

and

$$Var[\psi_{out}] = E[\psi_{out} - \overline{\psi}_{out}]^2. \quad (10)$$

The fundamental goal of optimum filters is to minimize the susceptibility of the estimated parameters to the non-deterministic components of the input signals. The variance of the output signal, a positive quantity given by Eq. (10), is indeed the parameter primarily minimized by the DPLMS method.

Furthermore, depending on the application, specific requirements in terms of the deterministic components of the filter output can sometimes also be desirable (e.g., finite length in time of the filter output response in order to achieve an accurate estimation of results even for piled-up signals, superimposed on the tails of previous events). This general class of constraints can be expressed as a separate functional of the filter coefficients:

$$\epsilon_n^2 = (\overline{\psi}_{out}(n) - \psi_0(n))^2. \quad (11)$$

All the single quadratic terms can be conveniently weighted and summed together to derive a global quadratic term:

$$\epsilon^2 = a_0 Var[\psi_{out}] + \sum_{k=1}^{K} a_k \epsilon_k^2 \quad (12)$$

which can be minimized against the entire set of the $N$ filter coefficients. The set of the resulting $N$ linear expressions represents a linear system, expressed in matrix form as:

$$A \cdot X = B \quad (13)$$

for which the column vector $X$, representing the optimum filter coefficients, can be straightforwardly derived.

2.1 Additional optimum filter constraints

To better understand the advantage of imposing additional constraints in the optimum filter synthesis procedure, let us neglect the fundamental goal of noise minimization, assuming noiseless filter input signals and represent the generic $k$-th radiation interaction process in the detector with the $\delta$-like function:

$$d_k(t) = E_k \cdot \delta(t - T_k). \quad (14)$$

This explicitly conveys the energy $E_k$ and arrival time $T_k$ information about the $k$-th event detected.

The optimum filter input signals can be considered as the result of the radiation interaction process in the experimental setup, thus expressed as the continuous-time convolution of the $d_k(t)$ term with the system impulse response function, here expressed with an event-dependent term $h_k(t)$, and an event-independent term $h(t)$.

$$\psi_{in,k}(t) = d_k(t) * h_k(t) * h(t). \quad (15)$$

The $h_k(t)$ function typically models the fundamental detection process, e.g., the current signal response to the $k$-th $\delta$-like unitary radiation event. This term may represent the current signal at the front-end readout electronics induced by the charge-collection in a direct-detection system (e.g., based on Germanium crystals) or the photo-detector current signal in an indirect-detection system based on scintillators. Since each radiation event undergoes a different conversion process (e.g., leading to a specific charge collection profile as a function of the interaction coordinates in large volume Ge detectors), the $h_{d,k}(t)$ functions are in principle all different from each other. In contrast, the $h(t)$ function typically models the voltage response of the subsequent electronic portion of the system to a unitary-area $\delta$-like current input signal (e.g., the front-end electronics, the interconnections, and the DAQ anti-aliasing filter). This term is here assumed as a time-invariant function, independent of the specific radiation event of interest.

The event-dependent impulse responses $h_{d,k}(t)$ are here assumed as non-negative functions, of constant area term (thus neglecting the detector intrinsic energy resolution limitation), and positively defined within a given time-interval $[0 - T_{max}]$, typically much shorter than the event-independent $h_S(t)$ impulse response function length. For example, in the case of solid-state detectors, the $h_{d,k}(t)$ functions, which are mainly modeling the detector charge collection profile, could typically last hundreds of nanoseconds. In contrast, the $h_S(t)$ function, modeling the charge-sensitive-preamplifier impulse response, could easily exceed hundreds of microseconds.

Additional contributions superimposed to the optimum filter input signal, independent of the radiation interaction in the detector, can be regarded as generated or picked up during the detector signals readout and processing phase. Let us then model these contributions with two separate terms: a time-dependent one, $n(t)$, which models the zero-average value system noise to be minimized by the optimum filter, and a constant one, i.e., $c$, which models any DC-offset possibly present.

$$\psi_{in,k}(t) = d_k(t) * h_k(t) * h(t) + n(t) + c. \quad (16)$$

Retrieving the energy and time information of the event of interest from the many-sample signal in Eq. (16) can be efficiently performed by applying proper FIR filters. As an example, energy estimation filters should provide results proportional to the $E_k$ parameter in Eq. (14) and ideally independent of other terms in Eq. (16). This well-defined, multi-objective goal can be easily implemented in the DPLMS optimum filter synthesis by specifying proper filter constraints, as discussed in the following.

## 2.2 Zero-area constraint

Among the filter properties expressed by Eq. (11), the zero-area constraint is especially useful to reject the baseline DC offset level of the experimentally acquired signal. The average value of the signals' baseline (i.e., without any radiation event) is usually a non-zero level. Consequently, a non-zero-area energy reconstruction filter would provide a non-zero output value (i.e., the so-called estimation "pedestal") when provided with baseline input signals. Therefore, all the subsequent energy estimations would also be shifted by the same amount. In the mathematical representation of the filter input signal of Eq. (16), the zero-area constraint translates into a filter response that is independent of the constant term $c$.

At least three different approaches can be used to deal with this issue. The simplest one, mainly used with peak-sensing ADCs, is based on: i) estimating the pedestal level during the system calibration phase and ii) compensating the subsequently computed pulse amplitude values by that constant value. However, this method is suboptimal because any long-term temperature drift or low-frequency pick-ups (e.g., 50 Hz disturbances) may alter the supposedly constant value of the pedestal, directly leading to estimation errors.

A second approach, popular with DAQ systems based on free-running ADCs, is based on repeatedly updating the estimated baseline value (e.g., by a digital filter or fitting algorithm) to provide a continuous correction of the parameters of interest over the entire acquisition phase. This technique may also be susceptible to the previously mentioned drawback due to the random and potentially long time interval between the baseline estimation and the event of interest. If needed, all of these can be implemented with two independently synthesized DPLMS optimum filters, independently estimating the signal baseline and the pulse amplitude values.

A third approach is based on a single zero-area FIR filter which directly disentangles the estimated values of the parameter of interest from any superimposed baseline non-zero value [12]. This technique, also implementable with DPLMS synthesized filters, usually provides better rejection of the low-frequency noise but may reduce the processing efficiency in the case of a higher rate of events due to the generally longer length of the FIR filter.

The zero-area filter FIR approach thus usually provides overall better estimation results, and it is often the solution of the choice in low-count-rate scenarios. However, in high-count-rate scenarios, the independent estimation of the signal baseline level may significantly increase the processing efficiency (e.g., equivalently decreasing the system dead-time) as opposed to the zero-area filters and be the preferable solution, despite its slightly worse estimation results.

We will derive the zero-area filter constraint with an applicative example. Due to the FIR filter's linearity and Eq. (9) (which shows that the output of an FIR filter is constant with a constant input), specifying the desired zero-area property of the optimum filter implies adding just a single punctual constraint, (e.g., setting $n = 0$ and a unitary input signal $\overline{\psi}_{in} = 1$).

$$\overline{\psi}_{out}(0) = \sum_{i=1}^{N} x_i \cdot \overline{\psi}_{in}(i - N) = \sum_{i=1}^{N} x_i. \quad (17)$$

This new punctual constraint, specified in the form of Eq. (11), shows that the quadratic functional effectively tends to zero with the filter area:

$$\epsilon^2_{(zeroarea)} = (\overline{\psi}_{out}(0) - \overline{\psi}_0(0))^2$$
$$= (\overline{\psi}_{out}(0) - 0)^2 = \left(\sum_{i=1}^{N} x_i\right)^2. \quad (18)$$

## 2.3 Pile-up-rejection constraint

Another helpful filter property that can be expressed by Eq. (11) is the rejection of the pulse "pileup" effect.

The pulse "pileup" condition is the superposition of the event-of-interest and random amplitude tails from the previous events. It is an experimental condition that may cause severe pulse-amplitude misestimations if not adequately addressed. This can be seen in Fig. 1 for simulated signals. For a given rate of randomly occurring events, the number of "piled-up" signals is likely to increase in the presence of slowly decaying pulse tails, such as the ones usu-ally added by preamplifier circuits and modeled by the $h(t)$ term in Eq. (16). In such a case, the long pulse tail is usually removed with a preliminary processing step (e.g., pole-zero compensation, moving-window deconvolution) before the noise-optimized shaping filter is applied. However, these techniques typically assume perfectly exponentially decaying pulse tails and might not always provide optimal performance depending on the experimental conditions. Moreover, apart from the self-adjusting systems, explicitly specifying the decay time value of the signal superimposed tails may be required.

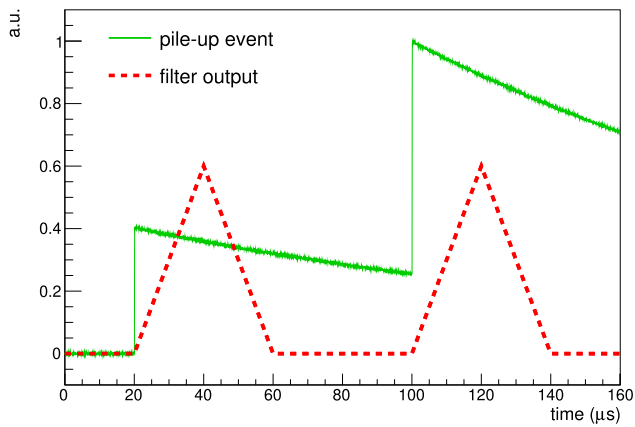

**Fig. 1** This figure shows two simulated germanium detector "piled-up" pulses at the DPLMS filter input (green continuous line) and the corresponding filter output signals (red dashed line), proving that the "pile-up" condition is removed at the filter output

Conversely, specifying the pileup rejection constraint for the DPLMS synthesized filters neither require any a priori information on the decay time value, nor does it assume purely exponentially decaying pulses. From the abstract viewpoint, optimally implementing the pulse "pile-up" constraint would entirely remove the $h(t)$ term in Eq. (16), i.e., like applying a $h_{dec}(t)$ filter such that:

$$h(t) * h_{dec}(t) = \delta(t). \tag{19}$$

The DPLMS quadratic term corresponding to the pile-up rejection constraint for generic shaped pulses may be expressed as:

$$\epsilon^2_{(pileup)} = \sum_k a_k \cdot \epsilon_k^2 = \sum_k a_k \cdot (\overline{\psi}_{out}(k))^2$$
$$= \sum_k a_k \cdot \left( \sum_{i=1}^N x_i \cdot \overline{\psi}_{in}(k - N + i) \right)^2. \tag{20}$$

Under a simplified assumption of purely exponentially decaying pulses, the Eq. (20) requires specifying a single punctual constraint (i.e., for a single $k$ value).

### 2.4 Flat-top constraint

Trapezoidal-shape weighting functions (WFs) are extensively used for estimating the energy released by radiation events in semiconductor detectors. Apart from their low-cost implementation by programmable logic devices [13], this WF class simultaneously achieves the two contrasting goals of rejecting i) the electronic series noise (best obtained with a triangular WF) and ii) the ballistic deficit effect (best obtained with a rectangular or, at least, a flat-topped WF). Moreover, in DAQ systems based on free-running ADCs, implementing a flat-topped WF of at least one (but practically a few) sampling intervals also compensates for the misestimation errors associated with the random occurrence in time of the detector pulses relative to the ADC sampling comb.

However, manually selecting the optimal flat-top length of the trapezoidal WF in a given experimental setup usually implies many time-consuming iterations to verify the relative effectiveness of various settings. In contrast, the new extension of the DPLMS algorithm automatically determines the optimal length and position of the WF flat-top in any given experimental setup.

This new option speeds up the optimum filter synthesis process and potentially improves the quality of the result by reducing the number of manually introduced constraints compared to the original DPLMS algorithm and the other methods in the literature.

For example, in [5] a two-step procedure is proposed, consisting of (i) automatically calculating the noise-optimized WF with no flat-top constraints and (ii) manually adding a flat-top section with the desired length at the center of the original WF. However, this procedure has two main drawbacks: (i) the final filter would consist of a manually altered WF, thus leading to a suboptimal performance, and (ii) a preliminary, independent step would be required to derive the original detector current signal, i.e., the $h_k(t)$ term in Eq. (16). Since the subsequent optimum filter synthesis algorithm assumes only $\delta$-like input signals, any additional term $h(t)$, modeling the filter input signal in Eq. (16), must be first identified and then explicitly removed.

The DPLMS algorithm overcomes those two drawbacks, not only by directly synthesizing an optimum filter with user-specified flat-top length and position for any signal shape but also by automatically synthesizing, at request, fully-featured optimum filters including flat-top length and position.

Here, we will review the conceptual assumptions behind the procedure to impose the flat-top constraints in the original DPLMS algorithm. According to Eq. (16), let us assume the detector-generated component of the filter input signal $\phi_{in}$, i.e., $d_k(t) * h_k(t)$, as completely deterministic, by supposing $h_k(t) = \delta(t)$, and a constant amplitude and fixed time relationship with the ADC sampling clock for the $d(t)$ term. Moreover, let us assume also that some signal-uncorrelated noise $n(t)$ is superimposed on the signal and that, for the sake of simplicity, $c = 0$. Such an ideal signal data set practically corresponds to the output signals generated by injecting a fast pulser signal of constant shape, amplitude, and time relationship with the ADC sampling clock into the readout electronics and DAQ system. Under these assumptions, the average filter input signal $\overline{\psi}_{in} = E[\phi_{in}]$ approximates the noiseless front-end electronics response to a $\delta$-like charge release in the detector.

Once the reference pulse signal $\overline{\psi}_{in}$ is calculated, the desired filter flat-top constraint can be implemented through Eqs. (9), (11) and (12), according to the procedure illustrated in [7].

For example, should a symmetrical WF with one sampling interval long flat-top be required and assuming $n_{trigger}$ as the beginning of the sampled reference pulse $\overline{\psi}_{in}$, the two values for $n$ in Eq. (9) would be $n = n_{trigger} + N/2$ and $n = n_{trigger} + N/2 + 1$. Additionally, the desired WF punctual values in Eq. (11), e.g., $\psi_0 = 1$ and the corresponding weighting coefficients, would be set in Eq. (12).

Two examples of DPLMS optimum filters, synthesized with and without the previously described flat-top constraint, are shown in Fig. 2.

This procedure, although effective in most cases, is unfortunately not suitable if the experimental condition prevents acquiring the emulated $\delta$-like charge release signal in the detector by injecting test pulses into the front-end electronics (e.g., the GERDA experiment at the Gran Sasso underground laboratories (LNGS) of the Italian Institute for Nuclear Physics (INFN) [14]).

In such cases, specifying the flat-top filter constraints can only rely on the experimental detector signals. However, since those signals usually have variable amplitudes and shapes, i.e., $h_k(t) \neq \delta(t)$ in Eq. (16), the previously mentioned procedure, based on the average pulse shape calculation, would not be fruitful.

To introduce the basic idea behind the first DPLMS extension and remark on how flat-topped filters effectively reduce the ballistic deficit effect in the energy estimations, let us start again from the detector signal model of Eq. (16). Because discussing the general case would require a comprehensive analysis beyond the scope of this paper, here is presented a simplified case with detector charge collection profiles $h_k(t)$ modeled by randomly delayed $\delta$-like functions, within a given $T_{max}$ long time interval:

$$h_k(t) = \delta(t - T_k) \qquad (21)$$

with $0 < T_k < T_{max}$.

Since the $A_k$ terms are in principle unknown, specifying the flat-top constraints for the $\psi_{in,k}$ noisy input signals as in Eq. (9) would be impossible. However, a new flat-top constraint expression, independent of the unknown $A_k$ terms, can be derived instead by following a different conceptual approach. First, two independent flat-top constraints expressions for each of the $\psi_{in,k}$ signal can be written:

$$\begin{cases} \psi_{out,k}(N/2) = \sum_{i=1}^{N} x_i \cdot \psi_{in,k}(i - N/2) = A_i \\ \psi_{out,k}(N/2 - 1) = \sum_{i=1}^{N} x_i \cdot \psi_{in,k}(i - N/2 - 1) = A_i. \end{cases} \qquad (22)$$

and then the two expressions of Eq. (22) can be combined:

$$\sum_{i=1}^{N} x_i \cdot \psi_{in,k}(i - N/2) = \sum_{i=1}^{N} x_i \cdot \psi_{in,k}(i - N/2 - 1)$$

$$\sum_{i=1}^{N} x_i \cdot (\psi_{in,k}(i - N/2) - \psi_{in,k}(i - N/2 - 1)) = 0$$

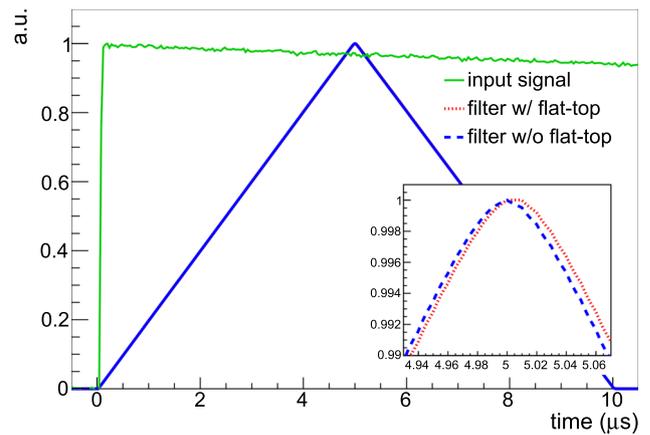

**Fig. 2** Two examples of optimum WFs, i.e., the convolution of the simulated, exponentially decaying DPLMS filter input signal (green, $\tau = 150\,\mu s$) and the two DPLMS FIR filters (not shown), synthesized in the case of simulated series noise. A triangular WF (blue) is generated without specifying any flat-top constraint, while a minimally flat-topped WF (one ADC sampling interval long, i.e., 10 ns, red) results instead from the procedure mentioned in the text

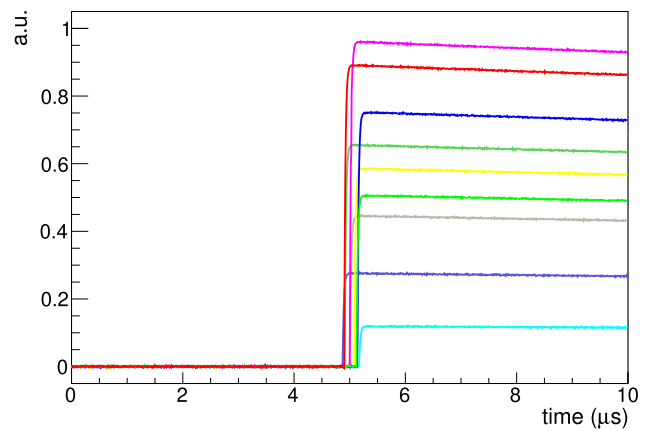

**Fig. 3** A subset of the simulated detector signals in the simplified case of Eq. (21), for $T_{max} = 500$ ns

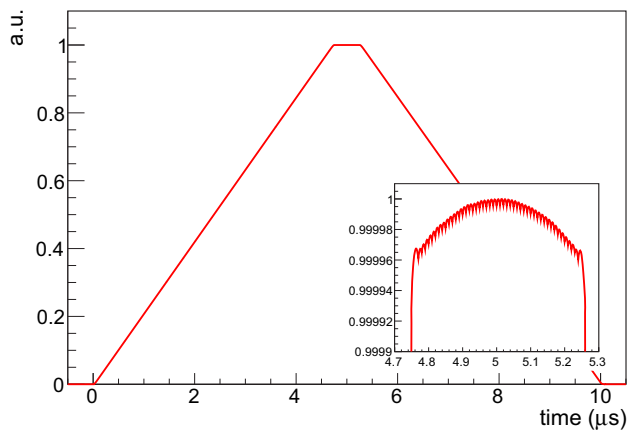

**Fig. 4** The WF corresponding to the optimum FIR filter synthesized in case of series noise, under the assumptions of Eqs. (16), (21) and $T_{max} = 500$ ns, showing that the automatically synthesized flat-top width matches the time distribution of the filter input signals

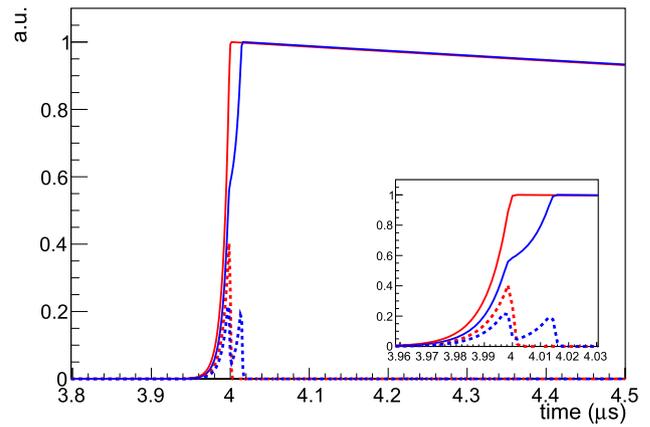

**Fig. 5** The two simulated BEGe detector SSE (red) and MSE (blue) noiseless signals at the preamplifier output and their associated current pulses (dashed lines)

$$\epsilon_k^2 = \left( \sum_{i=1}^{N} x_i \cdot \frac{d\psi_{in,k}(i - N/2)}{dn} \right)^2. \quad (23)$$

Figure 4 shows the WF corresponding to the optimum FIR filter synthesized in the case of series noise superimposed to the simulated input signals, partially shown in Fig. 3 and modeled according to Eqs. (16) and (21). The flat-top width matches the simulated charge collection time spread.

From the viewpoint of Eq. (16), estimating the energy released in the detector should provide a result proportional to the $A_k$ parameter and independent of any other term. Since it is the flat-top section of the WF that guarantees the independence of the estimated energy parameter from the $h_k(t)$ terms, using non-flat-topped WFs results in pulse height estimations instead of energy estimations (Fig. 4).

This is shown in Fig. 6, by comparing the result of applying a trapezoidal WF and a semi-gaussian WF to a basic data set of simulated BEGe detector's noiseless signals, represented in Fig. 5, one corresponding to a single-site event (SSE) of interaction within the detector and the other to a multi-site event (MSE). The two simulated events, although modeled by significantly different $h_k(t)$ terms in Eq. (16), do release precisely the same amount of energy $A_k$ in the detector.

The two WFs are scaled to provide unitary maximum pulse output with the SSE event, i.e., proportional to the energy released in the detector (Fig. 6). However, the MSE simulated pulse results in different pulse-height estimations with the two WFs: the trapezoidal WF provides the correct value in both cases, corresponding to the amount of deposited energy, while the semi-gaussian WF systematically underestimates it.

While flat-topped WFs are always a valuable feature in the case of non-instantaneous and non-deterministic charge collection profiles in the detector, they usually decrease the filter effectiveness from other perspectives (e.g., the electronic noise rejection). Before applying the DPLMS optimum filter synthesis, to achieve the best possible energy resolution, it can sometimes be helpful to apply an input signal alignment procedure to avoid synthesizing optimum filters with excessively long flat-tops. This specific issue will be practically addressed in Sect. 3.3 and, more thoroughly, in a forthcoming, dedicated paper, together with the discussion of the general case of Eq. (16).

Because the DPLMS filter synthesis is based on the blending of all the constraints discussed so far (e.g., those concerning the electronic noise, pulse pileup, ballistic charge deficit, input signal DC offset) and the ones considered only in [7] (e.g., expressed in the frequency domain), the optimum filter properties are a function of a set of weighting coefficients. These are usually tuned iteratively.

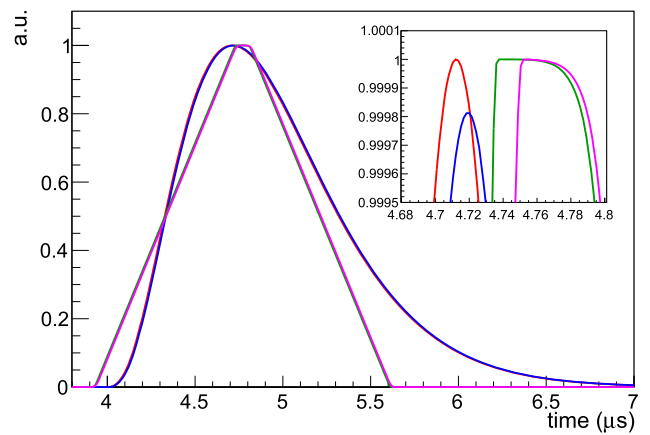

**Fig. 6** The shaping filter output signals from the two simulated BEGe detector signals in Fig. 5 are shown in this plot, with the SSE (MSE) output signal reported in red (blue) and green (violet) for the case of trapezoidal and semi-gaussian WFs, respectively

## 2.5 Optimum filter synthesis for multi-channel applications

In the case of multi-channel experiments, the DPLMS method not only allows synthesizing filters optimized for the noise auto-correlation of the individual channel of interest, but also for the inter-channel noise cross-correlation function. In Sect. 5, we introduce the multi-channel extension and show the improved results obtained with simulated signals. The underlying idea behind this approach is that, apart from the independent noise sources at the single device level (e.g., series and parallel noise), additional sources of noise (e.g., microphonics, pick-up electromagnetic disturbances) usually share a common physical origin, with a significant amount of cross-correlation between channels. A multidimensional optimum filter can significantly improve the precision of the estimation by considering both the channel noise auto-correlation and the noise cross-correlation between different channels.

## 3 Optimum filter synthesis with DPLMS method

The DPLMS method for optimum filter synthesis considers all the experimental noise components without assuming any underlying noise model, which usually leads to superior results compared to other methods [8]. This section of the paper will consider the properties of the FIR filters for energy estimation in various noise configurations and apply the first DPLMS extension to optimize the flat-top width of the filters automatically.

The first results presented in this section are obtained with simulated signals to appreciate the effects of the various issues. We simulated noiseless signals corresponding to the typical waveforms of High Purity Germanium (HPGe) detectors of about 1 MeV energy. In particular, we tuned the simulation parameters to be consistent with the front-end electronics and DAQ system of the GERDA experiment [14,15] (i.e., front-end electronics based on charge sensitive preamplifier with RC continuous reset with $C_F = 0.35$ pF, $R_F = 500$ M$\Omega$ and 14 bits flash analog-to-digital converter running at 100 MHz sampling rate with reduction to 25 MHz by summing up four consecutive samples and downsampling the sequence by a factor of four).

The noiseless signals were subsequently superimposed to variable intensity voltage (series) and current (parallel) electronic noise components and sinusoidal disturbances of varying amplitude and frequency.

Additionally, this section will report on the synthesis and application of the DPLMS filter in the case of a semi-coaxial HPGe detector (from Canberra Semiconductor), operated in the Milano Physics Department by the Nuclear Physics Group. This detector is in a standard, commercially available configuration, encased in a vacuum cryostat filled with liquid nitrogen and electrically connected to an integrated charge-sensitive preamplifier, internally equipped with a pole-zero compensation circuit.

In the following, we will illustrate in detail the procedure of optimum filter synthesis with the DPLMS method, explicitly focusing on the layout of the various matrices associated with the noise (Sect. 3.1), reference signal (Sect. 3.2), and flat-top (Sect. 3.3) constraints.

As already recalled, the DPLMS algorithm derives the optimum filter by globally minimizing the quadratic term $\epsilon^2$ of Eq. (12) with respect to the filter coefficients. Because, $\epsilon^2$ depends on the weighting coefficients of the constraints, so does the resulting optimum filter. Consequently, the optimum filter search usually consists of an iterative process. The weighting coefficients of the various constraints are progressively tuned towards the best overall result for the specific application of interest. Minimizing the computational burden associated with the DPLMS algorithm implementation is also a valuable aspect, briefly introduced in the continuation.

Each generic DPLMS optimum filter constraint, a quadratic function of the filter coefficients $x_i$, can be represented from a purely mathematical viewpoint by the following expression:

$$a \cdot \left( \sum_{i=1}^{N} x_i \cdot c_i - d \right)^2 \tag{24}$$

where $c_i$ and $d$ are generic constant terms and $a$ is the constraint weighting coefficient. Equation (24) can be equivalently represented in matrix form as:

$$a \cdot \left( X^T \cdot C \cdot C^T \cdot X - 2 \cdot C^T \cdot X + d^2 \right) \tag{25}$$

where $X$ and $C$ are column vectors of length $N$. Equation (25) shows that, by partially recollecting the filter coefficients terms $X$ among all the similar expressions, it is possible to combine all the constraints into a single equivalent quadratic form:

$$X^T \cdot M \cdot X - 2 \cdot V^T \cdot X + s \tag{26}$$

where $M$ is a symmetrical matrix, $V$ is a column vector and $s$ is a scalar.

The minimum value of this quadratic form, i.e., the point in the $N$-th dimensional space of the filter coefficients for which the first-order derivatives of the quadratic form are equal to zero, corresponds to the solution of the linear system:

$$M \cdot X = V. \tag{27}$$

Because of the two previously mentioned key points, the most computationally efficient way to derive the optimum filters with the DPLMS method, rather than combining the whole set of punctual constraints altogether, involves rearranging them into separate sub-groups based on their associated high-level intent, e.g., rejecting noise, pile-up, ballistic deficit, and subsequently assigning them a distinctive weighting coefficient. By doing this, five independent quadratic forms, as in Eq. (25), typically originate, each of them representing one of the main filter requirements to be satisfied, according to the associated weighting coefficients. Finally, thanks to the linearity property of the operator, each of the quadratic forms can be formally derived, independently of the weighting coefficient, to obtain a final linear system composed of five independently weighted elements:

$$\left(\sum_{i=1}^{N} a_i \cdot M_i\right) \cdot X = \sum_{i=1}^{N} V_i. \tag{28}$$

Because the only vector $V_i$ is different from zero, among those associated with the various filter constraints previously mentioned, it is the one corresponding to the unitary filter output. The final linear system of Eq. (13) can be expressed as:

$$A = a_1 \cdot NoiseM + a_2 \cdot RefM + a_3 \cdot ZeroAreaM$$
$$+ a_4 \cdot PileUpM + a_5 \cdot FlatTopM \tag{29}$$
$$B = a_2 \cdot RefV \tag{30}$$

Once the optimum filter $X$ is calculated, as a function of the $a_i$ weighting coefficients, the filter performance with respect to the various requirements can be evaluated. If required, a new iteration can be carried out with adjusted coefficients that cope better with the experimental condition, ranging from highly effective ($a_i = \infty$) to fully ineffective ($a_i = 0$).

### 3.1 Noise matrix

The matrix associated with noise minimization is built by summing multiple correlation matrices of experimental baseline waveforms (i.e., signals with no detector pulses), as:

$$NoiseM = baselineM^T \cdot baselineM \tag{31}$$

where the matrix $baselineM$ rows are the experimental baseline waveforms. Since the zero-area filter constraint independently accomplishes the input signal DC-offset rejection, the baseline signals used to build this matrix are shifted to have a zero average value. Three significant examples of noise matrices associated with some basic noise configurations are shown in Fig. 7.

Correlation matrices are always symmetrical, with the greatest values around the principal diagonal, as shown by the visual representations in Fig. 7, while their specific layout depends on the actual noise properties. The top-panel layout

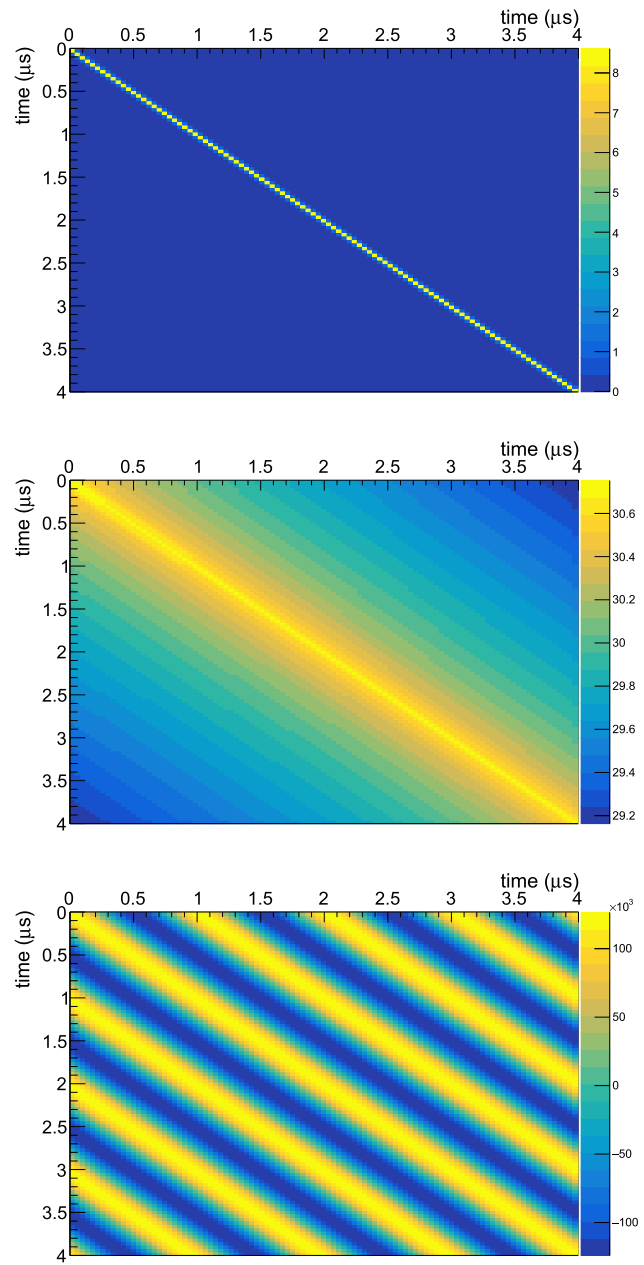

**Fig. 7** Noise matrix layouts in three configurations: (i) series noise (top panel); (ii) series and parallel noise (middle panel) and (iii) series noise and sinusoidal disturbances (bottom panel)

in Fig. 7 represents purely series noise, with absolutely predominant positive values in a very narrow band around the principal diagonal. The mid-panel layout of Fig. 7 represents purely parallel noise, this time showing slowly and monotonically decreasing values towards the two opposite matrix corners. The bottom panel layout of Fig. 7 represents the case of a small amount of series noise superimposed to a sinusoidal disturbance at 5 MHz frequency. The typical periodic correlation pattern of sinusoidal disturbances usually appears from the matrix layout, as in this case, also allow-ing a quick estimation of the fundamental frequency of the disturbance.

Apart from the specific examples illustrated here, we would like to remark that the correlation noise matrices can be not only a valuable means of synthesizing optimum filters out of the experimental signals but also an efficient tool for noise investigation and monitoring its consistency over time.

### 3.2 Reference signal

As discussed in Sect. 2, in a few cases, the deterministic input signal component is used to specify some optimum filter requirements. This signal can be straightforwardly derived from the input signals, which are averaged, amplitude normalized, and DC-offset compensated.

The associated terms of the reference matrix and vector are calculated as:

$$RefM = RefSignal^T \cdot RefSignal \tag{32}$$

and

$$RefV = RefSignal^T. \tag{33}$$

The decaying tail of the reference signal is also used to specify the goal of pile-up removal in the filter synthesis.

### 3.3 Flat-top matrix

The optimum filter flat-top feature is essential for reducing the ballistic deficit effect and obtaining a proper energy estimation of experimental signals collected from large volume detectors with non-instantaneous charge release. As previously recalled, a new extension of the DPLMS method (Sect. 2.4) straightforwardly copes with this issue by automatically synthesizing optimal filters of appropriate flat-top length. The layout of the associated "flat-top" constraint matrix starts from the time derivative of the detector pulse waveforms:

$$signalDer[i] = signal[i] - signal[i-1]. \tag{34}$$

Instead of proceeding with the same approach used for the "noise" matrix layout, in this case, it is equivalent and computationally more convenient to build first an intermediate matrix with the underived detector pulse waveforms and then convolve it with a simple 2D filter, represented by the $2 \times 2$ matrix:

$$M_{der} = \begin{pmatrix} 1 & -1 \\ -1 & 1 \end{pmatrix}. \tag{35}$$

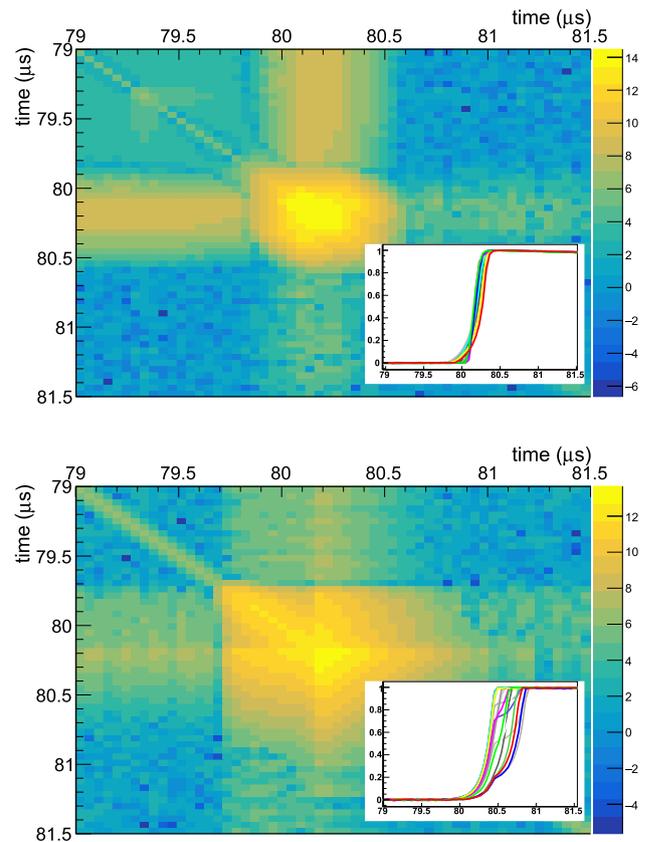

**Fig. 8** The non-trivial portions of two "flat-top" constraints matrices are shown, together with a few corresponding time signals in the insets

This matrix can be straightforwardly expressed with the equation:

$$FlatTopM = (signalM^T \cdot signalM) * M_{der} \tag{36}$$

where the $signalM$ rows contain the underived signals. The final single 2D convolution by $M_{der}$ provides the equivalent result of individually applying the time derivative to each of the waveforms.

Two "flat-top" constraints matrix examples are illustrated in Fig. 8. Only the central regions of the matrices, with values significantly different from zero, are shown. This corresponds to the rising edge regions of the detector signals reported in the insets.

The first example (top panel of Fig. 8) refers to signals collected in Milano, with a HPGe detector irradiated by a $^{60}$Co source, showing a nearly constant shape and a fast charge release. The second example (bottom panel) refers to simulated signals, roughly reproducing the case of a Broad Energy Germanium (BEGe) detector [16], i.e., a p-type HPGe detector by Canberra Semiconductor, previously used in the GERDA experiment because of its remarkable performance in terms of energy resolution and pulse shape discrimination of signal against the background.

Since the "flat-top" matrices originate from the time derivative of the detector signals, their highest data values are always associated with the rising edge portion of the waveforms. Moreover, detector signals with faster charge release transients, like those from the coaxial detector in the top panel, produce a more compact high data value section than the BEGe-like detector signals in the bottom panel, which have a slower and less uniform charge release.

Like in the case of the "noise" matrices, visually inspecting the "flat-top" matrices may also provide valuable hints about possible detectors' peculiarities or the trigger's efficiency.

Although outside the scope of this paper, it is also worth mentioning that with this extension, the flat-top synthesis is based on the variability of the detector charge release over time so that its position and length automatically provide the minimum weighted impact on the fulfillment of the other filter requirements.

### 3.4 Optimum filter synthesis

As already discussed in Sect. 2, the DPLMS filter is the solution of a linear system $A \cdot x = B$ (Eq. 13). The matrix $A$ and the vector $B$ are the linear combinations of the previously mentioned constraint matrices and vectors, with their weighting coefficients ($a_i$) appropriately tuned according to the experimental conditions.

Monitoring the filter efficiency, and satisfying the desired requirements, provides helpful information at the end of the synthesis procedure and the tuning phase of the weighting coefficients. The filter noise rejection capability can be represented by the output filter variance, is:

$$outNoise^2 = X^T \cdot NoiseM \cdot X \qquad (37)$$

where $X$ is the column vector of the filter coefficients and $NoiseM$ is the noise matrix.

### 4 Examples of optimum filter synthesis

To demonstrate the effectiveness of the DPLMS method with a variety of noise configurations, we will show some remarkable examples of optimum filter synthesis.

In Fig. 9 the results of optimum filter synthesis with simulated data in two different noise conditions are reported, showing the filter input signals as green lines, the optimum FIR filters as blue dotted lines, and the filter outputs signals, i.e., the convolution of the two previous signals, as red dashed lines. The optimum FIR filter shapes, as shown in Fig. 9, primarily depend on the input filter signal properties such as its deterministic form and the superimposed stochastic noise. It typically reflects the input signal fronts with a time-correlated gap in their shapes. On the other hand, the FIR filter output signal does not explicitly depend on the input filter signal shape since it represents the optimum weighting function for an ideal detector delta-like impulse. In order to show the effectiveness of the DPLMS method, we considered in these simulations two primary noise scenarios with already well-established optimum filter results in the literature. The top panel of Fig. 9 shows a pure series (voltage) noise, associated with an almost triangular DPLMS calculated overall weighting function. The bottom panel of Fig. 9 shows a combination of series and parallel (current) noise, associated with a narrow cusp-like DPLMS calculated overall weighting function. The two calculated results are well in agreement with the theoretical expectations, e.g., in [17], for a cusp-like optimum shaping filters. In the case of a combination of series and parallel noise, the cusp width depends on the ratio of the two noise contributions, as expected. It is wider (ultimately leading to a triangular shape) for prevalent series noise and narrower (ultimately resulting in a $\delta$-like function) for dominant parallel noise.

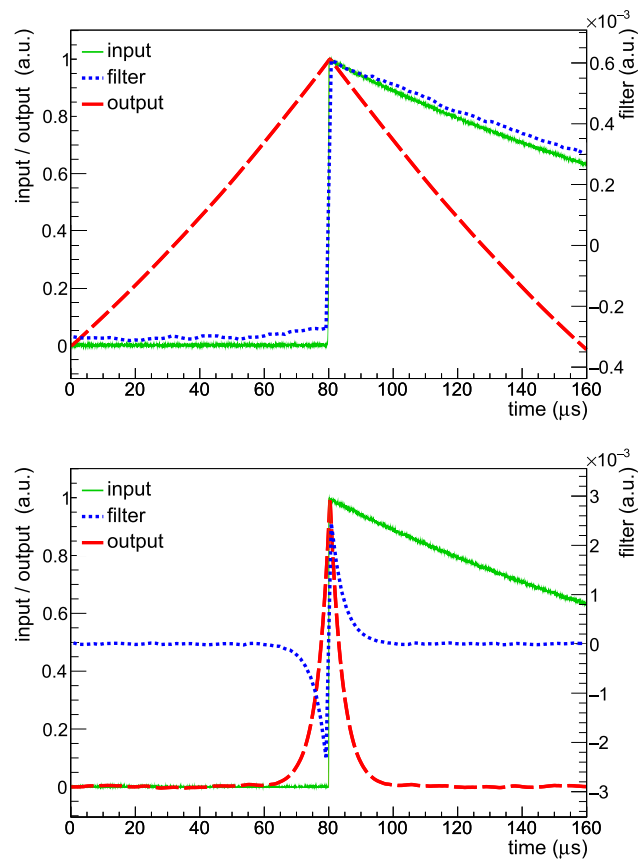

**Fig. 9** Filter synthesis results with simulated deterministic waveforms superimposed to series noise (top) and a combination of series and parallel noise (bottom), with input signal shown as green lines, optimum filters as blue dotted lines, and output signal as red dashed lines. Input and output signals, sharing the same vertical axis, are normalized to the unity (as for Figs. 10 and 11)

### 4.1 Optimum filters with sinusoidal disturbance

One of the significant advantages of the DPLMS-synthesized filters is their robustness against any sinusoidal disturbance superimposed on the signals of interest. In the following, we will show some optimum filter synthesis results with simulated deterministic waveforms, series noise, and sinusoidal disturbances of various frequencies. This includes 50 Hz, 20 kHz, and 1 MHz, thus spanning the whole expected range of conducted disturbances in the experimental environment.

In the first case (top panel of Fig. 10), with a disturbance repetition much longer than the optimum filter equivalent shaping time, the filter input signals appear almost homogeneously DC-shifted. Consequently, the synthesized optimum filter (blue dotted line) has nearly zero area, and the filter output signal (red dashed line) rejects the low-frequency noise components, thanks to its two negative symmetrical lobes. A very similar, empirically derived filtering technique provided valuable results in the GERDA experiment [18], where the non-standard operation in liquid Argon of the Ge detectors resulted in a considerable amount of low-frequency noise superimposed on the signals of interest.

In the more critical and hard-to-predict second scenario (middle panel of Fig. 10), the disturbance repetition of 50 $\mu$s is much closer to the optimum filter equivalent shaping time. Consequently, the synthesized optimum filter (blue dotted line) and the filter output signal (red dashed line) are significantly different from their ideal counterparts in the top panel of Fig. 9, derived for pure series noise, due to the oscillatory components associated with the disturbance rejection.

In the last case (bottom panel of Fig. 10), with a disturbance repetition of 1 $\mu$s (much shorter than the optimum filter equivalent shaping time) the synthesized optimum filter (blue dotted line) and the filter output signal (red dashed line) appear very similar to their ideal counterparts. Indeed, energy estimation filters generically provide some degree of intrinsic high-frequency rejection so that adding the specifically simulated disturbance did not significantly modify the optimum filter shape in this case.

A comparison of the energy estimation results obtained with the DPLMS optimum filters and a few commonly used sub-optimal filters are reported in Sect. 4.3.

### 4.2 Optimum filters with flat-top constraint

The results presented in the following derive from the two signals datasets previously described in Sect. 3.3, i.e. the experimental waveforms acquired from an HPGe detector operated at Milano University, and some simulated waveforms, reproducing BEGe detector signals [16]. The DPLMS filters synthesized for the two datasets have very different shapes. In the case of the HPGe detector (top panel of Fig. 11), the DPLMS synthesized filter, and the output filter have both a relatively short equivalent shaping time. In contrast, in the case of the BEGe-like signals, simulated with no parallel noise superimposed (bottom panel of Fig. 9), the DPLMS synthesized filter and the output filter have a much longer equivalent shaping time, almost corresponding to the ideal shapes expected in the case of pure series noise.

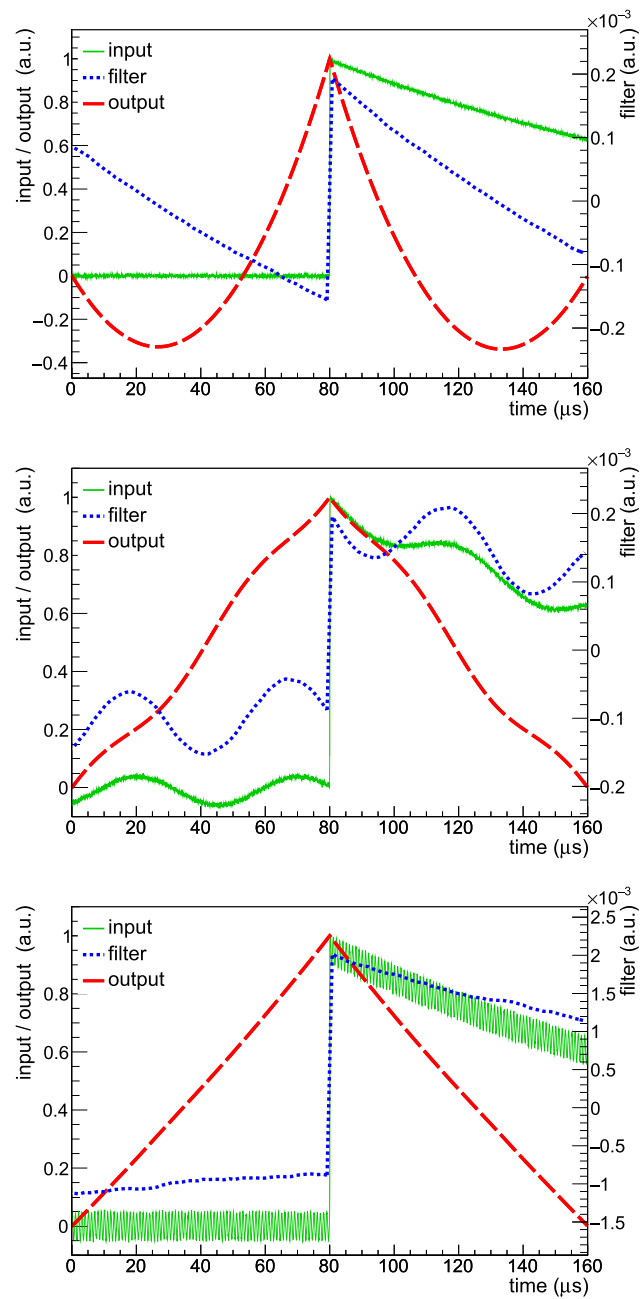

**Fig. 10** Filter synthesis results with simulated deterministic waveforms superimposed to series noise and time-uncorrelated sinusoidal disturbances at 50 Hz frequency (top panel), 20 kHz frequency (middle panel) 1 MHz frequency (bottom panel)

The insets of Fig. 11 show the magnified flat-top regions of the two optimum filters, each one superimposed to its shape of the single-site events, thus resulting in a shorter flat-top length and a better fulfillment of other filter constraints (e.g., electronic noise minimization).

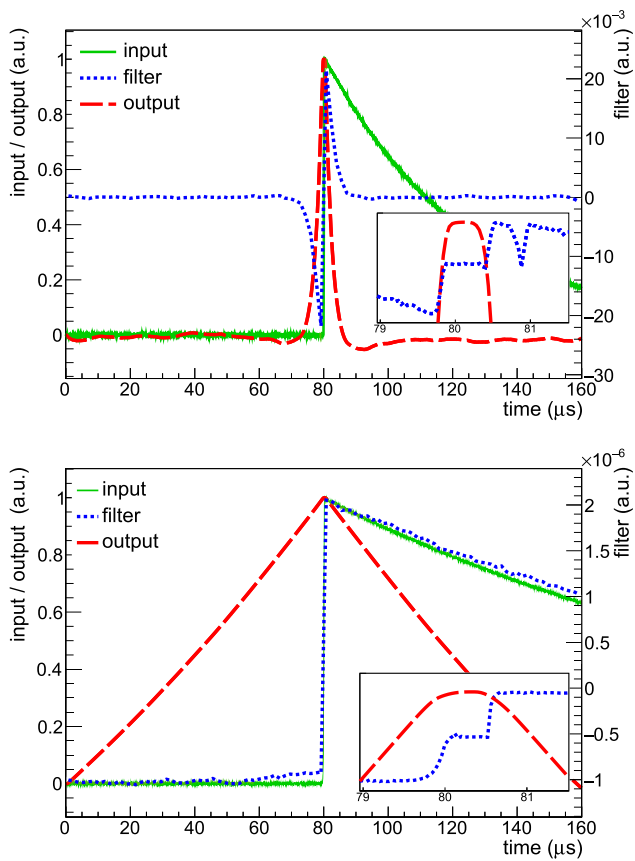

**Fig. 11** The results of filter synthesis with added flat-top constraints for the two cases considered in the text, with insets better showing the filter shapes around the flat-top regions (with vertical scales re-adjusted)

Following the previous consideration, it is also worth pointing out that in the case of offline signal processing, preliminarily re-aligning the acquired pulse waveforms may sometimes improve the effectiveness of the optimum filter synthesis. Indeed, the usually applied trigger algorithms, based on constant fraction discriminators or fixed threshold comparators, usually lead to a signal dataset that overestimates the filter flat-top length if compared with more refined triggering techniques, e.g., based on the charge collection profile of the events and, as a consequence, reduce the potential effectiveness of the optimum filter in satisfying other goals, e.g., the electronic noise rejection.

### 4.3 Comparison of performance

The figure-of-merit of a shaping filter for energy resolution, supposing input signals of constant energy, is directly related to the amount of variability of the output signal. The energy estimation is derived by sampling the filter output signal at a fixed time interval from the triggering event or extracting the maximum value in a reasonable time interval around it.

In the case of simulated waveforms with constant height, we defined the figure-of-merit of the energy estimators as the relative width of the distributions (i.e., the ratio of the full width at half-maximum of the lines to their centroids), with an ideal zero-value target. Conversely, in the case of the detector waveforms, we defined the figure-of-merit of the energy estimators as the absolute full-width at half-maximum of the two $^{60}$Co lines at 1173 keV and 1332 keV, with ideal target values given by the ideally expected energy resolution.

We selected two well-established and validated digital FIR filter alternatives in comparison with the DPLMS-synthesized filters. The first one implements a semi-gaussian shaping [19], which is usually available in the analog signal processing spectroscopy systems but can nonetheless be also digitally implemented, while the second one belongs to the "ZAC" filter class, i.e., the cusp-like, zero-area filters used in the GERDA experiment [18].

Since a DPLMS synthesized filter is intrinsically optimized with respect to the actual signal and noise characteristics, we first selected also the best-performing ZAC filter, by tuning the various filter parameters, or degrees of freedom in the filter class, to identify the best performing combination. However, since we also wanted to quantify the potential improvements of the noise-optimized filtering techniques, compared to the non-adaptive, fixed filtering techniques, we selected just one semi-gaussian filter of fixed shaping time, corresponding to the one initially selected in GERDA.

As expected, the DPLMS-synthesized optimum filters provide the best energy estimation results in all cases.

corresponding output signal, with the flat-top lengths automatically adjusted by setting the appropriate constraint in the DPLMS filter synthesis. It is worth mentioning that, while the flat-top regions of the filters in Fig. 11 are about the same length for both cases, the comparison of the corresponding flat-top matrices in Fig. 8 highlights a larger number of significantly non-zero values in the case of the simulated BEGe detector signals. Although beyond the scope of this paper, it is nonetheless interesting to briefly comment on this result by recalling that flat-topped filters improve the energy estimation quality mainly by reducing the ballistic deficit effect, which is, in turn, determined by the spread of the charge collection profiles of the pulses on an event-by-event basis. The BEGe-like signals considered in the paper were naively simulated, assuming that all the single-site interactions in the detector resulted in the same charge collection profile and that the multi-site interactions were composed of the previously considered single-site interactions with random amplitude and time distribution. Consequently, the synthesized flat-top length correctly reflected the time variability of the simulated multi-site events and not the deterministic

**Table 1** The performance of the three filters (relative energy resolution FWHM in ppm)

| Noise configurations | DPLMS  | ZAC    | Semi-gaussian |
|----------------------|--------|--------|---------------|
| Series               | 247(3) | 496(5) | 476(5)        |
| Series, parallel     | 566(5) | 660(7) | 663(7)        |
| Series, 50 Hz dist.  | 482(5) | 495(5) | –             |
| Series, 20 kHz dist. | 255(3) | –      | –             |
| Series, 1 MHz dist.  | 250(3) | 561(6) | 478(5)        |

The analytical optimum filter theory [17] states that the semi-gaussian and the ZAC filter class should provide good energy estimation results for any combination of series and parallel noise, superimposed to the signals of interest, provided the selected equivalent shaping time is appropriately matched to the longer or shorter values in the case of predominant series or parallel noise respectively.

On the other hand, the two filter classes perform very differently in the presence of low-frequency noise disturbances (like the 50 Hz sinusoidal disturbance assumed here). The ZAC filter class, thanks to its symmetrical, zero-area shape still provides excellent results. This is not the case for the semi-gaussian filter class, which would require an additional baseline subtraction step (like an analog baseline restorer circuit or a digital baseline fitting algorithm).

The energy estimation performance with middle and high-frequency narrow-band noise confirms the effectiveness of the DPLMS-synthesized filters, which completely reject the added disturbance and retrieve the same results as in the case of series noise alone. Conversely, while in the third case almost acceptable results can be directly achieved by the filters from the semi-gaussian and the ZAC classes, thanks to the general low-pass frequency nature of all the energy estimation filters, none of those filters provides enough rejection of the middle-frequency disturbance to produce reasonable energy spectra.

Table 1 reports the energy resolution in terms of part-per-million (ppm) relative full width at half maximum (FWHM) obtained with the DPLMS filters and the two benchmark filters.

To better investigate how a different amount of noise reflects into the energy resolution results obtainable with the three filtering techniques, we simulated a dataset of deterministic signals with an increasing amount of superimposed series and parallel noise and reported the corresponding energy estimation results in Fig. 12, in terms of the relative FWHM of the histogram lines.

Finally, we report on the energy estimation results obtained from the HPGe detector in Milano. The optimum filter synthesized with the DPLMS method (shown in the top panel of Fig. 11) results in an absolute FWHM value of $1.90 \pm 0.02$ keV and $1.99 \pm 0.02$ keV for the 1173 keV

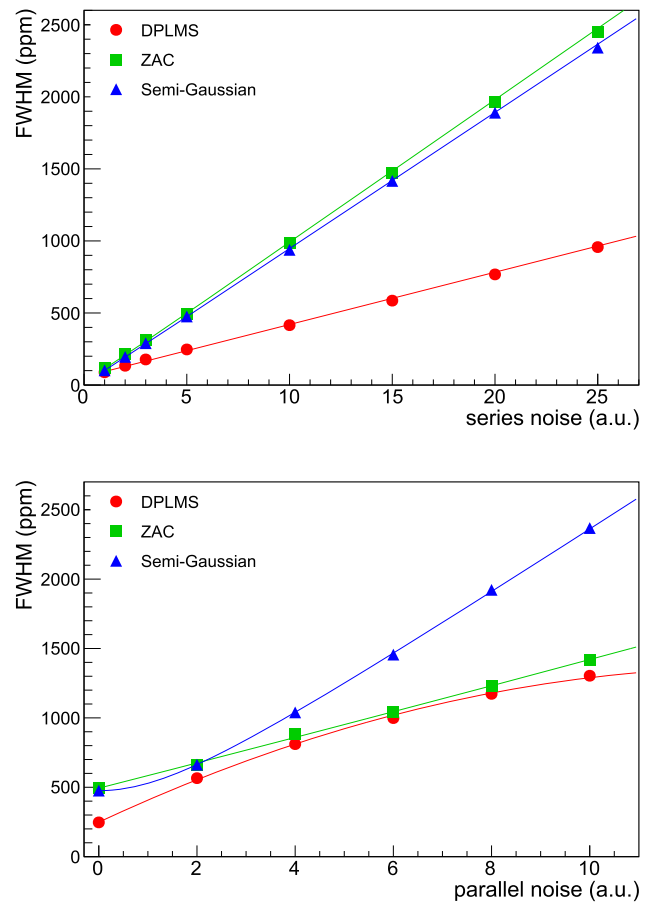

**Fig. 12** The energy estimation results (in terms of peak relative FWHM) obtained with the three filtering techniques: the DPLMS filters (red circles), the ZAC filters (green squares), and the semi-gaussian shaping filters (blue triangles) for an increasing amount of series noise (top panel) and of parallel noise (bottom panel)

and 1332 keV peaks, respectively. The corresponding results obtained with the best-performing semi-gaussian shaping filter are instead $2.25 \pm 0.03$ keV and $2.29 \pm 0.03$ keV, which is roughly 14% worse than the optimum filter. The results obtained with the best-performing ZAC filter are worse than 2.5 keV, suggesting that a symmetrical, zero-area filter shape, very different to the optimal filter shape synthesized by the DPLMS algorithm, does not really fit the experimental data.

To summarize, the previously reported results confirm that it is usually possible to achieve good energy resolution in spectroscopy applications even without direct optimum filter synthesis, provided adequate knowledge of the experimental noise conditions is available and a two-step procedure is followed. Namely: (i) a suitable filter class (e.g., cusp-like, ZAC, semi-gaussian) must be wisely selected among those commonly used or described in the literature and (ii) the filter parameters within that filter class (e.g., equivalent shaping time, flat-top length) that best match the actual noise conditions must be selected, usually by iterative testing.

In this respect, direct optimum filter synthesis methods, such as the DPLMS, clearly offer two significant advantages over the parametric iterative search of the best performing filter in a class of filters selected a priori. Firstly, the filter design phase is usually accelerated, which is especially important in multi-channel experiments and applications requiring subsequent filter readjustment during their long data-taking periods, and, secondly, an improvement of the energy estimation results is usually achievable, thanks to the many more degrees of freedom available.

Conversely, supposing the experimental noise configuration also includes non-fundamental, narrow-band harmonic disturbances, the only feasible option to derive effective filters would be with synthesis algorithms based on the experimentally acquired waveforms, such as the DPLMS algorithm, as no analytical calculation can deal with the complexity of the real-world noise scenarios.

## 5 Optimum filter synthesis with correlated noise

As discussed in Sect. 2.5, a new DPLMS extension allows further improvements of the energy estimation in multi-channel applications by exploiting the additional knowledge on the inter-channel noise correlation. This section will illustrate the application of the new extension with a simulated dual-channel system, although the same procedure can be easily generalized to any number of channels.

### 5.1 Correlation noise matrix and filter synthesis

The starting point of the DPLMS extension is the same as the standard method presented in Sect. 3. The information on the particular noise superimposed on the signals is extracted by a set of experimental baseline traces. In addition to the traces of the channel that we want to analyze (referred to as the first channel), in this case, the method requires a set of baseline traces from a second channel. If the two traces have a correlated noise, the method automatically takes advantage of this and produces a better energy estimation. On the other hand, if the correlation between the channels is zero, the method is equivalent to the original one.

Similar to the single noise matrix of Eq. (31), we calculate here a matrix with the input noise by combining the baseline traces of the two channels:

$$NoiseMatrixCorr = BaselinesCorr^T \cdot BaselinesCorr \tag{38}$$

where $BaselinesCorr$ contains the two baseline traces for each event; noting that here, the size of the correlated noise matrix is doubled compared to the noise matrix of Eq. (31).

A new DPLMS matrix can be calculated by summing all of the desired contributions:

$$A_{corr} = a_1 \cdot NoiseMatrixCorr + a_2 \cdot RefMatrix + \cdots \tag{39}$$

where the reference matrix and other additional constraints are the same as the single-channel method discussed in Sect. 3, including the information of the first channel alone.

The optimum filter can then be calculated by solving a linear system with the matrix of Eq. (39):

$$A_{corr} \cdot x_{corr} = B \tag{40}$$

with the known term $B$ fixed to the reference signal of the first channel. The number of linear equations of the system of Eq. (40) is twice the number of signal samples, and the same is true for the number of coefficients of the correlation filter $x_{corr}$. Considering the criteria adopted to synthesize $x_{corr}$, the first half of the filter has to be applied to the first channel, while the second half is for the correlated channel. Figure 13 reports two examples of the correlation DPLMS optimum filter synthesis: the blue dotted line is the optimum single-channel filter (same filters of Fig. 9), the green line is the correlated filter for the first channel and the red dashed line is the correlated filter for the second channel.

A first application of the method has been performed on simulated signals with the superposition of series noise only and a second channel with 70% correlated noise (top panel of Fig. 9). The three filters report the same shape and a different amplitude. Parallel noise has been included in the filter synthesis with a 70% correlated noise for the second channel as well. The resulting filters are reported in the bottom plots of Fig. 9. The single-channel filter has the typical narrow shape obtained with parallel noise (see Sect. 3), but this is not the case for the correlated filters (green and red lines) that have a shape typical of pure series noise. The explanation of this behavior is that the parallel noise is taken into account by the channel correlation, while the filters deal with the remaining series noise.

The described extension of the DPLMS method can be easily generalized to a multi-channel case to take advantage of the correlation noise between all the channels of the system by building a correlation noise matrix with all the baseline traces, as in the two-channel case of Eq. (38). Consequently, the DPLMS matrix and the linear system have a dimension that increases with the number of included channels. In practical applications, it can be useful to calculate a priori the noise correlation between channels and include in the filter synthesis only those with significant correlation.

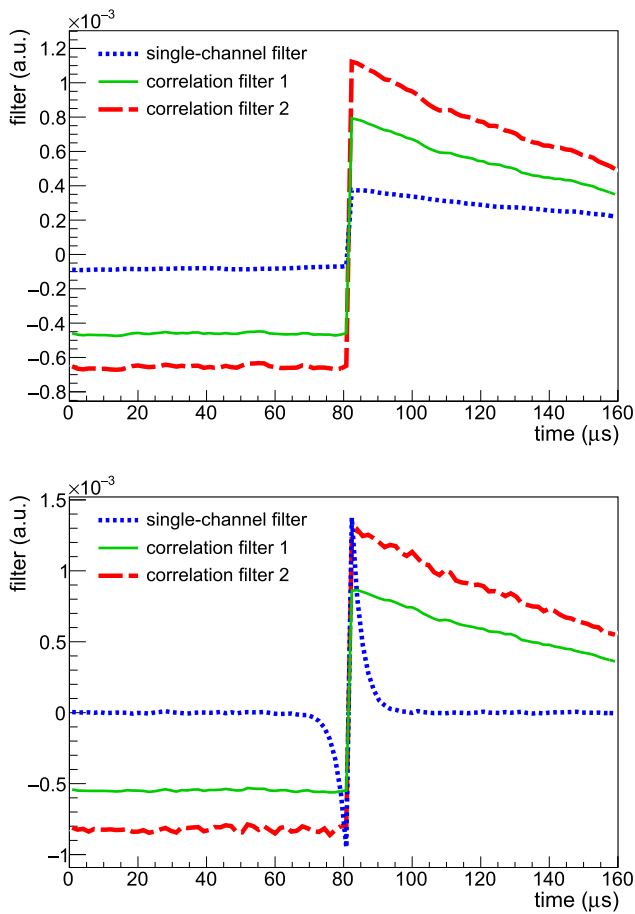

**Fig. 13** DPLMS optimum filter synthesis with correlated noise between two simulated channels. The blue dashed line is the single-channel filter, the green continuous line is the correlated filter for the first channel and the red long-dashed line is the correlated filter for the second channel

**Table 2** Results of the application of the DPLMS method extension with correlated noise

| Correlation | Series noise FWHM [ppm] | Parallel noise FWHM [ppm] |
| --- | --- | --- |
| 100% | 157(2) | 93(1) |
| 70% | 173(2) | 96(1) |
| 40% | 180(2) | 98(1) |
| 10% | 202(2) | 145(1) |
| 0 | 207(2) | 1848(18) |

### 5.2 Performance of correlated filters

The main purpose of the DPLMS method extension is to take advantage of channel correlation to reduce the effect of the noise, beyond the limit of the single-channel method. In order to show how the noise correlation can be exploited, we studied the performance, in terms of energy resolution, as a function of the correlation with a second channel. The results obtained spanning from full to zero correlation cases with intermediate steps are reported in Table 2. Two noise configurations are investigated: series noise only and the addition of a dominant parallel noise.

For the series noise configuration, the FWHM values of Table 2 show substantial resolution improvements of 13–25% with a strong correlation between the two channels ($\geq 40\%$), but also in the low correlation case of 10% we see an improvement of few percent. In the parallel noise configuration, the efficiency of the method is more noticeable with improvements larger than 90% in all cases: the channel correlation is able to almost remove the parallel noise and the width of the reconstructed peak is due to the remaining series noise.

The validity of the DPLMS method is underlined by the presented extension that exploits the noise correlation between two or more channels, and this is particularly aimed at multi-channel experiment applications.

## 6 Conclusions

The DPLMS method allows to automatically synthesize FIR filters taking into consideration the complex nature of the specific field of application, by globally satisfying a weighted combination of the possibly conflicting goals. These include the fundamental goal of noise minimization together with additional requirements, e.g., the zero-area constraint for non-zero baseline level removal and the pile-up rejection. Therefore, the DPLMS is a practical and effective tool to extract parameters of interest in radiation detectors, such as the deposited energy. Filters synthesized with the DPLMS method have been applied to simulated signals and to a real configuration with a germanium detector. An extension of the original DPLMS algorithm allows to create a filter with a flat-top of proper length, resulting in an enhanced energy resolution with respect to other algorithms. In complex experimental multi-detector systems, the extension of the DPLMS method can take advantage of the correlated noise between the channels to significantly improve the overall performance.

ted use, you will need to obtain permission directly from the copyright holder. To view a copy of this licence, visit http://creativecommons.org/licenses/by/4.0/.

Funded by SCOAP$^3$. SCOAP$^3$ supports the goals of the International Year of Basic Sciences for Sustainable Development.